\documentclass{skaox2006}
\begin{document}
   \title{Equatorial Imaging with e-MERLIN Including the Chilbolton Antenna}
 
   \titlerunning{Equatorial Imaging with e-MERLIN + Chilbolton}
   \authorrunning{Ian Heywood, Hans-Rainer Kl\"{o}ckner, Steve Rawlings}
 
   \author{Ian Heywood, Hans-Rainer Kl\"{o}ckner and Steve Rawlings}

   \institute{Oxford Astrophysics, Denys-Wilkinson Building, Keble Road, Oxford, OX1 3RH}

   \abstract{We discuss the equatorial imaging benefits that arise from the addition of the 25-metre dish at Chilbolton to the e-MERLIN array.
   Its inclusion considerably enhances the capabilities of e-MERLIN on and below the equator. This will become particularly important
   in the era of ALMA and other upcoming southern hemisphere facilities. We present simulated observations of point sources in the
   equatorial region of the sky which is the target area for many existing sky surveys. We find that the additional baselines created by the inclusion of the Chilbolton
   dish favourably adjust the beam shape of e-MERLIN to a more compact and circular shape, with significantly reduced sidelobe structure. 
   Putting aside the benefits of increased collecting area, the modified
   beam shape has implications for more rapidly reaching a given completeness limit for equatorial surveys.}

   \maketitle
%
%
\section{Introduction}

The facilities at Chilbolton Observatory in Hampshire, UK, include a fully steerable 25-metre parabolic antenna which is mainly used for 
meteorological Doppler-polarisation radar. The use of this antenna in the MERLIN array (Thomasson, 1986) has been discussed
for many years, and with the provision of a fast fibre link it would become a prime candidate for inclusion in the 
e-MERLIN\footnote{\tt{http://www.merlin.ac.uk/e-merlin/}} array.

\begin{figure}
   \centering
   \vspace{215pt}
   \includegraphics{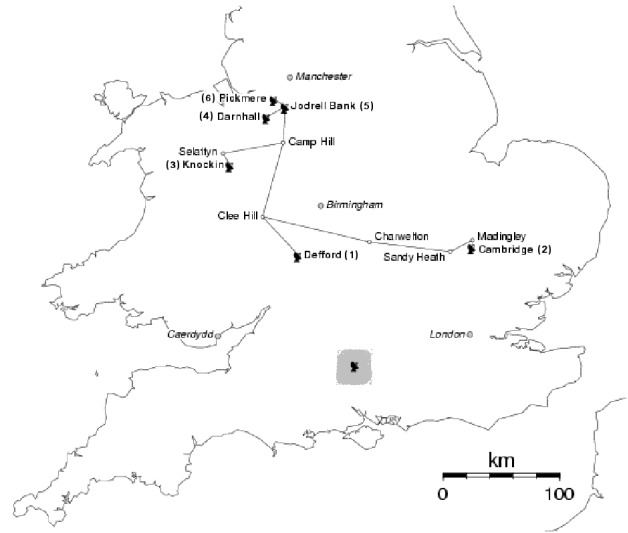}
   \caption{Map showing the current MERLIN array, with the location of the Chilbolton Observatory highlighted in grey.
            \label{fig:merlin}
           }
    \end{figure}

Figure~\ref{fig:merlin} is a version of the diagram presented on the online MERLIN user
guide\footnote{\tt{http://www.merlin.ac.uk/user\_guide/OnlineMUG/}} which has been modified in order to show the location of the Chilbolton antenna in addition to the locations of the existing MERLIN stations.
Antenna numbers for the existing MERLIN stations have also been added.
As can be seen, inclusion of the Chilbolton antenna will boost the number of long and intermediate-length baselines, as well as extending the north-south span of the array,
facilitating the sampling of a much greater range of spatial frequencies. This is particularly crucial for `snapshot' mode observations, and large-scale
survey programmes where on-source time may necessarily be rather brief.

In this article we demonstrate the improved $uv$-plane sampling of an e-MERLIN+Chilbolton array for both snapshot
and full synthesis equatorial observations. We present simulations demonstrating the response of both arrays to an unpolarized point source, and discuss the implications
of the inclusion of the Chilbolton antenna when carrying out equatorial surveys.

All observations are simulated using AIPS with standard imaging procedures. All simulations are at L-band, using 100 frequency channels of 
5 MHz each to simulate a 1.3 - 1.8 GHz contiguous band, assuming a spectrally flat source. Each antenna has an efficiency of 0.8 and system temperatures 
in the range 20 - 33~K. The 76-metre 
Lovell telescope is not included in the simulated array.

\section{Increased $uv$-plane coverage}

\begin{figure*}
   \centering
   \vspace{196pt}
   \includegraphics{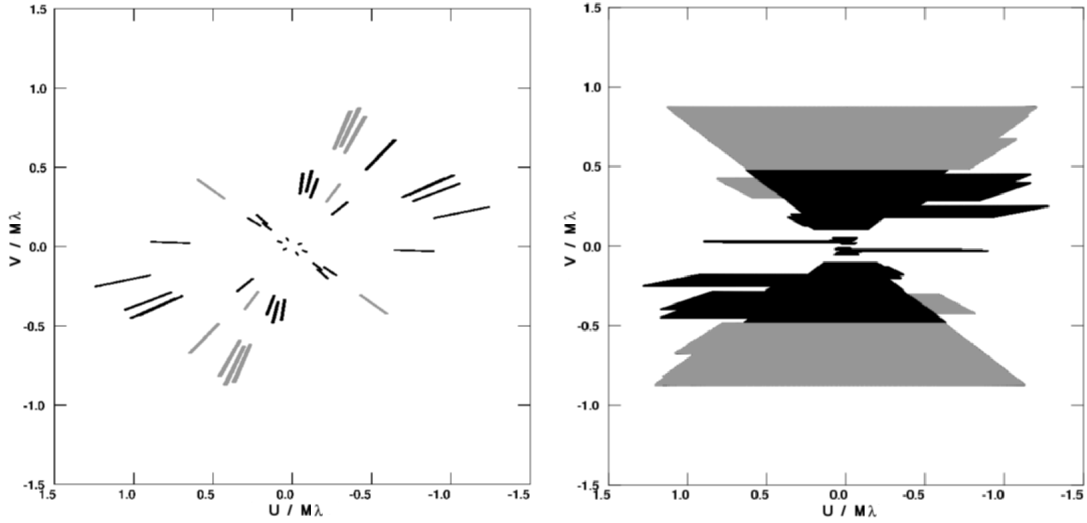}
   \caption{The $uv$-plane coverages for a snapshot (left) and 12-hour (right) observation of a sky patch at RA = 0 and Dec = 0 with an hour angle range of -6 to +6. The additional visibilities due to the 
   inclusion of the Chilbolton antenna are shown in grey.
            \label{fig:uv_plots}
           }
\end{figure*}

Figure~\ref{fig:uv_plots} shows the $uv$-plane coverage for the e-MERLIN array plus the Chilbolton antenna for a 5-minute
snapshot observation (left panel) and a 12-hour track (right panel). The additional visibilities arising from the inclusion of the Chilbolton antenna are shown
in light grey. 

Each of these simulations is an observation of an equatorial source (with a Right Ascension and Declination of zero). The snapshot observation has an on-source time
of 5 minutes at an hour angle of zero and the 12-hour track has an hour angle range of -6 to +6. The latter scenario gives the best possible $uv$-coverage for a source 
at this position due to the opacity of the earth. The integration time assumed in both cases is 60 seconds. 

The key feature to note in both of these plots is that the additional Chilbolton baselines greatly reduce the
east-west bias of the array and increase the number of long and intermediate-length baselines. The image-plane implications of this modified
$uv$-plane coverage are demonstrated in Section 3. 

\section{Imaging considerations}

\begin{figure*}
   \centering
   \vspace{196pt}
   \includegraphics{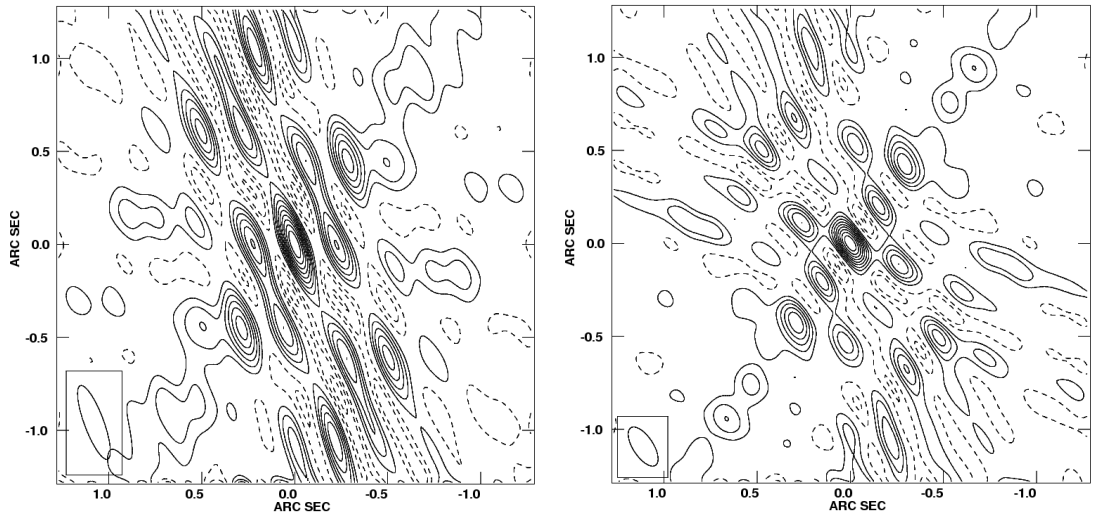}
   \caption{Simulated, zero-noise dirty images of a 5-minute snapshot observation of a 1~mJy point source at RA = 0, Dec = 0 and hour angle = 0, with e-MERLIN 
   without (left) and with 
   (right) the Chilbolton antenna. These images are equivalent to the dirty beams of the two arrays. Contour levels are -30\%, -20\%, -10\%, 10\%, 20\%, ... ,90\%, 100\% times the peak
   in the map.
            \label{fig:beams}
           }
\end{figure*}

Simulated, zero-noise dirty images of a 5 minute snapshot observation
of an equatorial 1-mJy point source are presented in
Fig.~\ref{fig:beams}. These images correspond to a 6-antenna e-MERLIN
array without (left) and with (right) the Chilbolton antenna,
observing with the 1.3-1.8~GHz contiguous band. Since these are dirty
images and, since they are observations of a point source, they
represent the point spread function of the array. A Gaussian fitted to
the half power level of the central lobe is plotted in the lower left
of each panel. The dimensions of these ellipses in arcseconds are 0.42
$\times$ 0.10 arcsec$^{2}$ (position angle = 20.4$^{o}$, without
Chilbolton) and 0.24 $\times$ 0.10 arcsec$^{2}$ (position angle =
33.3$^{o}$, with Chilbolton).

The benefits of the additional baselines to the Chilbolton station are immediately apparent. The dominant, near vertical structure and strength of the sidelobes
in the beam pattern are significantly reduced with the addition of the Chilbolton antenna. The central lobe of the beam is also much closer to circularity.

\begin{figure*}
   \centering
   \vspace{196pt}
   \includegraphics{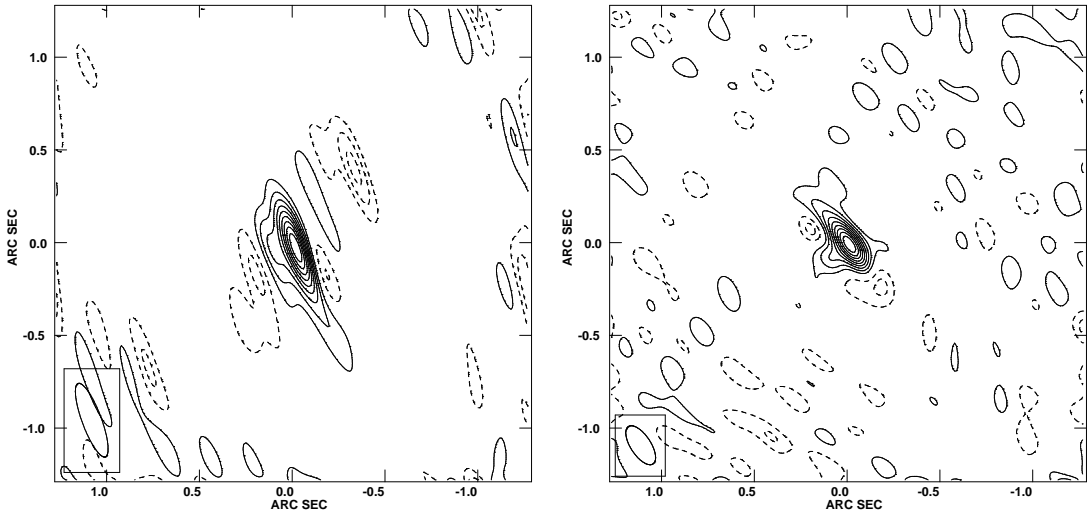}
   \caption{The result of cleaning simulated 5 minute snapshot images with e-MERLIN with Gaussian noise included. The image with the Chilbolton antenna included is on the right. Contour levels are -30\%, -20\%, -10\%, 10\%, 20\%, ... ,90\%, 100\% times the peak
   in the map (refer to Table \ref{tab:stats}).
            \label{fig:snapshots}
           }
\end{figure*}

\begin{figure*}
   \centering
   \vspace{196pt}
   \includegraphics{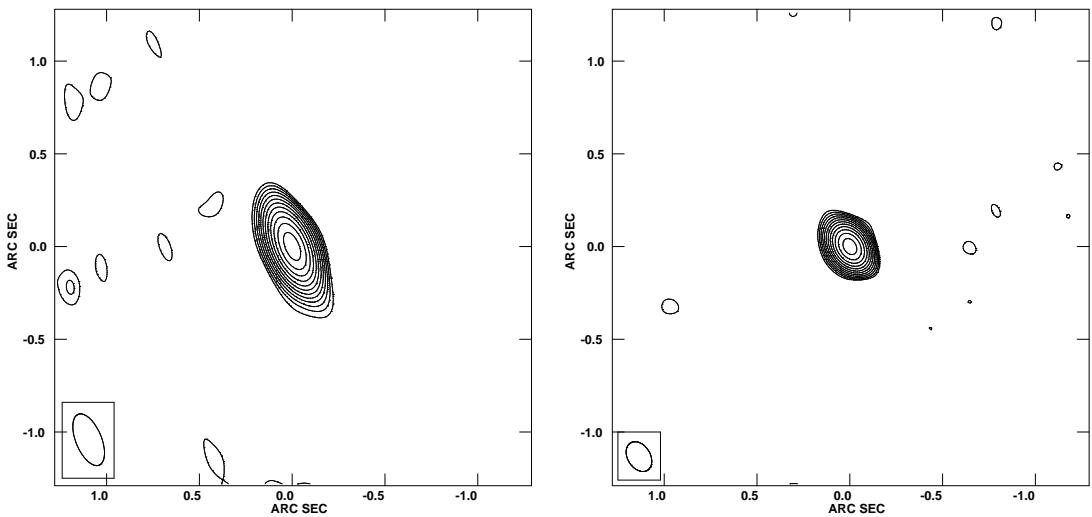}
   \caption{Cleaned images of a model point source with the simulated observation being a full 12 hour synthesis using e-MERLIN. The simulation with the Chilbolton antenna included is on the right. Contour levels
   are (--$\sqrt{2}$, 1, $\sqrt{2}$, 2, 2$\sqrt{2}$, 4, 4$\sqrt{2}$, 8, 8$\sqrt{2}$...) $\times$ $\sigma$, where $\sigma$ is the root mean square of the background noise (refer to Table \ref{tab:stats}).
            \label{fig:12hours}
           }
\end{figure*}

Figure \ref{fig:snapshots} shows cleaned, simulated snapshot images without (left) and with (right) Chilbolton, with Gaussian noise added, the level
of which is calculated by the AIPS task UVCON according to the antenna and observation parameters. These simulations obviously assume perfect phase and
amplitude calibration which is never the case in practice. Similarly, Fig.~\ref{fig:12hours} shows cleaned
images corresponding to a 12-hour track. The contour levels are adjusted for these images in order to display the $\sim\mu$Jy-level background noise.
Note that phase and amplitude errors would yield additional symmetric and anti-symmetric (relative to the peak) noise patterns, similar to those
seen in Fig.~\ref{fig:beams}.

Relevant parameters for these four images are listed in Table~\ref{tab:stats}. The values and uncertainties in the peak fluxes and integrated flux densities
are determined by fitting Gaussians to the centre of the image using the AIPS task IMFIT. The background RMS values in the image are returned by isolating 
the background region using TVWIN and executing the AIPS
verb IMSTAT. 

Note that in the case of Fig.~\ref{fig:snapshots} there is no improvement at all in the RMS achieved despite the increased collecting area. The decrease in RMS expected 
because of the $\sim$10\% increase in collecting area is masked by sources of systematic error.

\begin{table*}
\centering
\begin{minipage}{158mm}
\caption{Cleaned image parameters for Fig.~\ref{fig:snapshots} and Fig.~\ref{fig:12hours}, recalling that the input model is a 1-mJy point source. 
B$_{maj}$ and B$_{min}$ are the major and minor axes respectively of the ellipse fitted to the central lobe
of the beam at the half-power level. BPA is the position angle of this ellipse. $eM$ in the simulation column indicates use of solely the six e-MERLIN antennas whereas
$eM+C$ means that the Chilbolton antenna is added. The RMS increase due to the increases collecting area is apparent when comparing the values corresponding 
to the 12-hour tracks. \label{tab:stats}}
\begin{tabular}{lllllll}
\hline
Simulation        & Peak flux           & Integrated flux     & Background RMS & B$_{maj}$  & B$_{min}$ & BPA   \\
		  & (mJy/beam)          & density (mJy)       & (mJy/beam)      & (arcsec)  & (arcsec)  & (deg) \\
\hline
Snapshot ($eM$)   & 0.903~$\pm$~0.044 & 1.081~$\pm$~0.085 & 0.056         & 0.422  & 0.104   & 20.4\\
Snapshot ($eM+C$) & 0.936~$\pm$~0.063 & 1.024~$\pm$~0.115 & 0.058         & 0.241  & 0.108   & 33.3\\
12 hour ($eM$)    & 1.006~$\pm$~0.007 & 1.025~$\pm$~0.012 & 0.009         & 0.297  & 0.138   & 22.5\\
12 hour ($eM+C$)  & 1.013~$\pm$~0.007 & 1.009~$\pm$~0.011 & 0.007         & 0.174  & 0.122   & 31.9\\

\hline
\end{tabular}
\end{minipage}
\end{table*}

Assuming that the sidelobe structure can be successfully removed by whatever means, then when an extended Gaussian-like radio source is observed its brightness 
distribution is essentially convolved with the central lobe of the beam. 
This convolution will `smear' the flux density out over an area depending on the size of the source and the size of the beam, the result being another Gaussian\footnote{The 
convolution of any two Gaussians is another Gaussian, 
since \emph{(i)} the product of any two co-centred Gaussians is a Gaussian, \emph{(ii)} 
the convolution of any two functions is equal to the Fourier transform of their product, and \emph{(iii)} the Fourier transform of a Gaussian is another Gaussian.}. 
Given that a radio source will generally have a fixed brightness over the course of an observation, the observed peak brightness (i.e. the `height' of the resulting Gaussian) 
will be dependent on the size of the central lobe. Using a more compact beam, as in the case where the Chilbolton antenna is 
included, reduces the broadening and increases the peak brightness, allowing a flux limit to be reached faster in certain cases.

MERLIN observations of the Hubble deep and flanking fields have shown that a typical micro-Jansky radio source
has a characteristic size of $\sim$0.6 arcsec (Muxlow et al., 2005). The major axes of the snapshot beams are $\sim$0.4
arcseconds (without Chilbolton) and $\sim$0.2 arcseconds (with Chilbolton). 

With the above considerations in mind, we present quantitative estimates of the increased mapping speed provided by the inclusion of the Chilbolton antenna 
in Table~\ref{tab:maptimes}. 
These values have been calculated by convolving two-dimensional
Gaussians, measuring the difference in the peak value for the two arrays, and assuming the sensitivity has an inverse dependence on the square root
of the on-source time. Two models consistent with the full-synthesis beam shapes with and without Chilbolton, and two source models with major axes
of 0.6 and 0.3 arcsec (with minor axes half this value) are used. The calculation is performed with both parallel and orthogonal alignment of the major axes of the source
and beam, representing the best and worst case respectively for a random distribution of source alignments.

\begin{table}
\centering
\caption{Increase in mapping speed due to the inclusion of the Chilbolton antenna, expressed as a percentage of the observing time 
required for full synthesis observations using the regular e-MERLIN array. This is presented for sources with major axes of 0.6 and 0.3 arcseconds, aligned both parallel and orthogonal to
the major axis of the beam. \label{tab:maptimes}}
\begin{tabular}{ll}
\hline
Simulation & Factor \\
\hline
0.6 arcsec, aligned	& 85\% \\
0.6 arcsec, orthogonal	& 68\% \\
0.3 arcsec, aligned	& 64\% \\
0.3 arcsec, orthogonal	& 46\% \\
\hline
\end{tabular}
\end{table}

\section{Conclusions}

The geographical location of the Chilbolton antenna introduces baselines which complement the existing e-MERLIN array. This
is particularly evident in the case of equatorial observations, where the dominant east-west layout of the existing array biases the $uv$ coverage, resulting
in strong linear structure in the sidelobes of a synthesised beam which is highly eccentric.

Putting aside the increased sensitivity due to the $\sim$10\% increase in collecting area with the additional antenna, 
the synthesised beam is much more compact and circular for equatorial observations, facilitating more efficient detection
of characteristic $\mu$Jy radio sources, reducing the mapping speed by up to a factor of two (see Table~\ref{tab:maptimes} for favourably oriented sources.

The addition of the Chilbolton antenna would significantly enhance the power of 
e-MERLIN. There are many experiments that would benefit significantly from this enhancement, such as high-resolution
radio surveys complementing optical and infrared deep-fields such as those to be undertaken by VISTA,
and high-resolution radio observations of galactic and extra-galactic radio sources targetted with ALMA. Both e-MERLIN
and ALMA naturally complement each other due to their similar sub-arcsecond resolutions, despite their operating frequencies
differing by a factor of $\sim$100.

\begin{acknowledgements}
The authors would like to thank Simon Garrington of Jodrell Bank Observatory and Ken Craig of Rutherford Appleton Laboratory.
\end{acknowledgements}

\end{document}